\documentclass[twocolumn,english,aps,prb,twocolum,superscriptaddress,bibnotes,amsmath,amssymb,floatfix]{revtex4-1}
\usepackage[colorlinks=true,citecolor=blue,linkcolor=magenta]{hyperref}

\usepackage[markup=nocolor, authormarkupposition=left]{changes} 
\usepackage{soul}
\usepackage[utf8]{inputenc}
\usepackage[english]{babel}
\usepackage{amsmath,amsfonts,amssymb}
\usepackage[T1]{fontenc}
\usepackage{url}

\usepackage{amsmath}
\usepackage{amsfonts}
\usepackage{amssymb}

\usepackage{epstopdf}
\usepackage{graphicx}
\graphicspath{{./Figures/}}

\begin{document}
\title{Laser soliton microcombs on silicon}

\author{Chao Xiang}
\thanks{These authors contributed equally to this work.}
\affiliation{ECE Department, University of California Santa Barbara, Santa Barbara, CA 93106, USA}

\author{Junqiu Liu}
\thanks{These authors contributed equally to this work.}
\affiliation{Institute of Physics, Swiss Federal Institute of Technology Lausanne (EPFL), CH-1015 Lausanne, Switzerland}

\author{Joel Guo}
\affiliation{ECE Department, University of California Santa Barbara, Santa Barbara, CA 93106, USA}

\author{Lin Chang}
\affiliation{ECE Department, University of California Santa Barbara, Santa Barbara, CA 93106, USA}

\author{Rui Ning Wang}
\affiliation{Institute of Physics, Swiss Federal Institute of Technology Lausanne (EPFL), CH-1015 Lausanne, Switzerland}

\author{Wenle Weng}
\affiliation{Institute of Physics, Swiss Federal Institute of Technology Lausanne (EPFL), CH-1015 Lausanne, Switzerland}

\author{Jonathan Peters}
\affiliation{ECE Department, University of California Santa Barbara, Santa Barbara, CA 93106, USA}

\author{Weiqiang Xie}
\affiliation{ECE Department, University of California Santa Barbara, Santa Barbara, CA 93106, USA}

\author{Zeyu Zhang}
\affiliation{ECE Department, University of California Santa Barbara, Santa Barbara, CA 93106, USA}

\author{Johann Riemensberger}
\affiliation{Institute of Physics, Swiss Federal Institute of Technology Lausanne (EPFL), CH-1015 Lausanne, Switzerland}

\author{Jennifer Selvidge}
\affiliation{Materials Department, University of California Santa Barbara, Santa Barbara, CA 93106, USA}

\author{Tobias J. Kippenberg}
\email[]{tobias.kippenberg@epfl.ch}
\affiliation{Institute of Physics, Swiss Federal Institute of Technology Lausanne (EPFL), CH-1015 Lausanne, Switzerland}

\author{John E. Bowers}
\email[]{bowers@ece.ucsb.edu}
\affiliation{ECE Department, University of California Santa Barbara, Santa Barbara, CA 93106, USA}
\affiliation{Materials Department, University of California Santa Barbara, Santa Barbara, CA 93106, USA}

\maketitle

\noindent\textbf{ Silicon photonics enables wafer-scale integration of optical functionalities on chip. 
A silicon-based laser frequency combs could significantly expand the applications of silicon photonics, by providing integrated sources of mutually coherent laser lines for terabit-per-second transceivers, parallel coherent LiDAR, or photonics-assisted signal processing. 
Here, we report on heterogeneously integrated laser soliton microcombs combining both InP/Si semiconductor lasers and ultralow-loss Si$_3$N$_4$ microresonators on monolithic Si substrate. 
Thousands of devices are produced from a single wafer using standard CMOS techniques. 
Using on-chip electrical control of the microcomb-laser relative optical phase, these devices can output single-soliton microcombs with 100 GHz repetition rate. 
Our approach paves the way for large-volume, low-cost manufacturing of chip-based frequency combs for next-generation high-capacity transceivers, datacenters, space and mobile platforms.
}
\begin{figure*}[t!]
\centering
\includegraphics{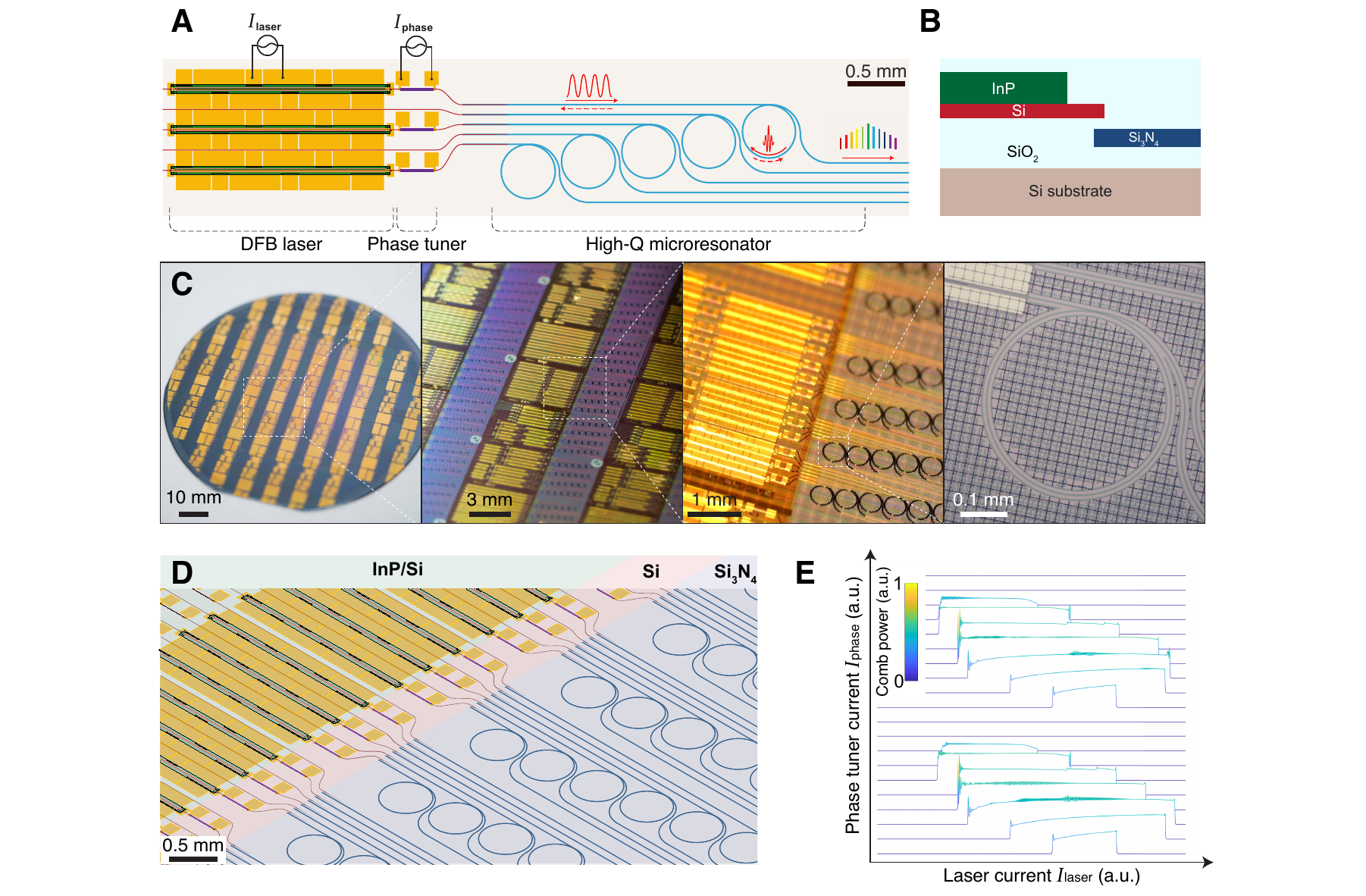}
\caption{\textbf{Device schematics, images and operation principle}. 
A. Design layout of a laser soliton microcomb device consisting DFB lasers, phase tuners and high-$Q$ microresonators on a common substrate. 
A continuous-wave (CW) signal (solid red line) emitted from the laser is partially back-scattered in the microresonator. 
The back-scattered signal (dashed red line) is sent back to the laser, triggering self-injection locking that assists the formation of soliton pulse stream inside the microresonator. 
The optimization of laser self-injection locking is realized by controlling the laser current $I_\text{laser}$ and phase tuner current $I_\text{phase}$.
B. Simplified device cross-section schematics. 
The laser is based on InP/Si. 
The microresonator is based on Si$_3$N$_4$. 
Silicon layer is used to deliver light from the InP/Si layer to the Si$_3$N$_4$ layer. 
C. Photographs showing the completed 100-mm-diameter wafer, zoom-in of multiple laser soliton microcomb dies, and a microscopic image showing a Si$_3$N$_4$ microring resonator with Si/ Si$_3$N$_4$ interface. 
D. Bird-view schematic illustration of the large-scale integrated laser soliton microcomb devices enabled by multilayer heterogeneous integration. 
The lasers, phase tuners and high-$Q$ microresonators are built on InP/Si layer, Si layer and Si$_3$N$_4$ layer, respectively.
E. Schematic illustration of the microcomb generation with sweeping laser current $I_\text{laser}$ and varying phase tuner current $I_\text{phase}$ (thus to vary the optical phase difference between the forward and backward signals), in the case of laser self-injection locking. 
This waterfall plot is based on a full nonlinear simulation shown in the Supplementary Information.
}
\label{Fig:1}
\end{figure*}
%%%%%%%%%%%%%%%%%%%%%%%%%%%%%%%%%%%%%%%%%%%%%%%%%%%%%%%%%%%%%%%%%%%%%%

Optical frequency combs (OFC) \cite{Udem:02, Cundiff:03} have revolutionized timing, spectroscopy and metrology\cite{Fortier:19, diddams_optical_2020}. 
Historically, OFCs have been made using femtosecond mode-locked lasers with supercontinuum generation for octave-spanning spectra, required for self-referencing. 
Discovered a decade ago \cite{DelHaye:07}, OFCs can also be generated in driven Kerr-nonlinear optical microresonators. 
These types of OFCs are commonly referred to as "microcombs". 
Microcombs can be operated in regimes where they form coherent, temporal, dissipative structures \cite{Lugiato:87, Akhmediev:95}, i.e. bright dissipative Kerr solitons (DKS) \cite{Herr:14}. 
DKS have unlocked the full potential of microcombs by providing coherent, broadband OFCs with repetition rates in the terahertz to microwave domains, and have been successfully employed in many system-level applications such as RF photonics\cite{Torres-Company:14, Wu:18}, coherent communication~\cite{Marin-Palomo:17, Corcoran:20}, astronomical spectrometer calibration~\cite{Suh:19, Obrzud:19}, massively parallel coherent LiDAR \cite{Riemensberger:20}, optical frequency synthesizers~\cite{Spencer:18} and photonic neromorphic computing \cite{Feldmann:21, Xu:21}. 
On a fundamental level, DKS microcombs have allowed explorations of a plethora of novel nonlinear dynamics and phenomena \cite{Xue:15, Yang:17b}. 

In parallel, there has been substantial progress in photonic integrated platforms \cite{Kovach:20} for microcomb generation, which to date include Si$_3$N$_4$\cite{Moss:13, Xuan:16, Ji:17, Liu:20b, Ye:19b}, AlN\cite{Jung:14, LiuX:20}, LiNbO$_3$\cite{He:19, Wang:19, Gong:20}, AlGaAs\cite{Pu:16, Chang:20, Moille:20}, GaP\cite{Wilson:20}, and SiC\cite{Lukin:20}. 
Among them, the leading platform is Si$_3$N$_4$, which has already been used widely in CMOS microelectronics as diffusion barriers and etch masks.
In addition to its native properties such as the absence of two-photon absorption, high Kerr nonlinearity, and weak Raman and Brillouin gains \cite{Gyger:20}, recent advances in fabrication of nonlinear Si$_3$N$_4$ photonic integrated circuits (PIC) have enabled optical propagation losses below 1 dB/m \cite{Ji:17, Liu:20b}. 
These ultralow losses have significantly reduced the soliton formation threshold power to the levels that integrated lasers can provide \cite{HuangD:19, McKinzie:21}, and have yielded soliton repetition rates in the microwave X-band \cite{Liu:20}.  
In combination with negligible thermal effects and almost purely Kerr-dominated frequency-dependent response \cite{Liu:20b}, reliable soliton generation can be attained without complex or fast laser tuning. 
Additionally, laser self-injection locking \cite{Liang:15, Pavlov:18} and hybrid integration of soliton microcombs with RSOAs \cite{Stern:18} or DFB lasers \cite{Shen:20} allow for current-initiated and electrically controlled modules with low electrical power \cite{Stern:18, Raja:19, Shen:20, jin2021hertz}.
 
A long-standing goal is to monolithically integrate lasers and high-$Q$ nonlinear microresonators onto a common silicon wafer.  
Heterogeneously integrated silicon photonics \cite{Komljenovic:16,liang_recent_nodate,zhang_07_uTP} offers a compelling solution using low-cost, industry-standard silicon substrates, aided by mature CMOS-compatible fabrication facilities.
High-performance, large-scale heterogeneous PICs with complete functionalities, including lasers \cite{fang_electrically_2006, liang_recent_2010, Xiang:20}, modulators \cite{xu_micrometre-scale_2005, reed_silicon_2010, hiraki_heterogeneously_2017}, and photodetectors \cite{xie_high-power_2016, Michel:10}, are disrupting optical interconnect technology and other applications \cite{roelkens_iii-vsilicon_2010, jones_heterogeneously_2019,sun_large-scale_2013}. 
The material-by-design nature enabled by heterogeneous integration offers best-in-class performance.
However, heterogeneous integration of high-power, narrow-linewidth, semiconductor lasers with high-$Q$ Si$_3$N$_4$ microresonators has not been demonstrated yet, because multiple material platforms (Si, Si$_3$N$_4$, and III-V) with significantly different optical properties have to be deployed and combined. 
Here we overcome these challenges and present the first demonstration of heterogeneously integrated laser soliton microcombs. 
This is achieved using ultralow-loss Si$_3$N$_4$ PICs based on the photonic Damascene fabrication process in conjunction with direct SOI wafer bonding and heterogeneous III-V integration. 
The result is a wafer-scale fabrication process that produces thousands of devices from one silicon wafer. 
Each device allows electrical initiation and control of soliton microcombs.
Our technology demonstrates the viability of large-volume, low-cost manufacturing of chip-based OFCs, and allows incorporation of soliton microcombs as new building block into existing complex silicon photonics systems, and in applications requiring reliable performance, small footprint, low cost and low power consumption, such as space or mobile platforms. 

%%%%%%%%%%%%%%%%%%%%%%%%%%%%%%%%%%%%%%%%%%%%%%%%%%%%%%%%%%%%%%%%%%%%%%
\begin{figure*}[t!]
\centering
\includegraphics{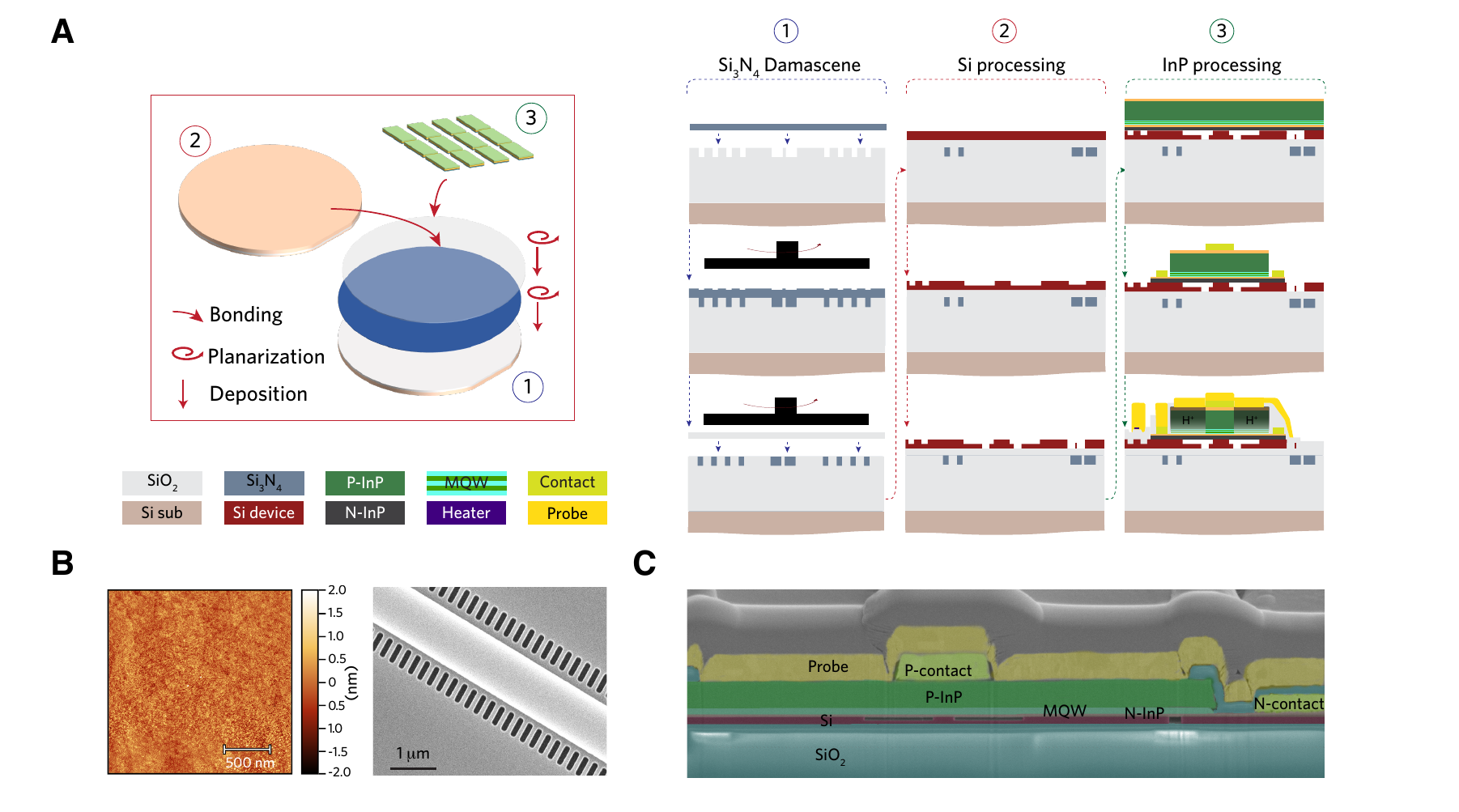}
\caption{\textbf{Simplified wafer-scale fabrication process flow}. 
A. Multilayer heterogeneous integration process. 
Left panel shows that a pre-patterned Si$_3$N$_4$ Damascene substrate is planarized and subsequently bonded with an SOI wafer and an InP MQW epi wafer. 
The wafer undergoes 1.) Si$_3$N$_4$ photonic Damascene process, 2.) Si processing and 3.) InP processing.
Right panel shows selected key steps in the wafer fabrication with 1.) Si$_3$N$_4$ photonic Damascene process including LPCVD Si$_3$N$_4$ deposition on the patterned SiO$_2$ substrate (top), excess Si$_3$N$_4$ removal and planarization (middle), and SiO$_2$ spacer deposition, densification and polishing (bottom); 
2.) Si processing including Si substrate removal (top), Si waveguide etch for laser and thermal tuner (middle), and for Si grating and Si-Si$_3$N$_4$ mode conversion taper (bottom); 
3.) InP processing including InP substrate removal (top), InP mesa etch (middle), and excess Si removal, laser passivation, contact formation, vias open, heaters, and probe metal formation (bottom).
B. Left: Atomic-force microscopy (AFM) shows 0.27 nm RMS roughness of the Si$_3$N$_4$ wafer surface, after the second CMP on the SiO$_2$ spacer, before bonding with the SOI wafer. 
Right: SEM image of Si grating for laser before bonded with InP MQW wafer.
C. False-colored SEM image showing the cross-section of the complete device at laser area with multilayer device structure.
}
\label{Fig:2}
\end{figure*}
%%%%%%%%%%%%%%%%%%%%%%%%%%%%%%%%%%%%%%%%%%%%%%%%%%%%%%%%%%%%%%%%%%%%%%

\textbf{Device schematic}. 
As shown in Fig. \ref{Fig:1}A, a chip-scale laser frequency comb consists of three main parts: a distributed feedback (DFB) laser, a thermo-optic phase tuner, and a high-$Q$ nonlinear microresonator, by leveraging multilayer heterogeneous integration\cite{Xiang:20} (Fig. \ref{Fig:1}B).  
Figure \ref{Fig:1}C shows the photographs of the complete wafer (before dicing into chips), multiple dies, single-die-level devices, and an optical microscope image of a Si$_3$N$_4$ microring resonator interfaced with Si layer.
As illustrated in Fig. \ref{Fig:1}D, the DFB lasers, phase tuners, and nonlinear microresonators are built on indium phosphide/silicon (InP/Si), silicon (Si), and Si$_3$N$_4$ layers, respectively. 
This vertical, multilayer structure is realized through sequential wafer bonding of a silicon-on-insulator (SOI) wafer and an InP multiple-quantum-well (InP MQW) epi wafer to a pre-patterned Si$_3$N$_4$ substrate that is fabricated using the photonic Damascene process \cite{Liu:20b,Xiang:20}. 
Deep ultra-violet (DUV) stepper lithography is used to ensure pattern alignment on each individual layer with an accuracy better than 100 nm. 
Here we directly apply heterogeneous integration on 100-mm-diameter Si substrates and process the entire substrate on the wafer scale. 
Our process could be further scaled up with larger wafer substrates such that it is compatible with standard CMOS foundry production lines \cite{jones_heterogeneously_2019}.

The DFB laser is a high-power, single longitudinal-mode pump source with an excellent side mode suppression ratio (SMSR) such that the laser wavelength can be tuned to a microring resonance\cite{coldren_diode_2012}. 
The laser has a 1.8-mm-long InP/Si gain section, and the grating is etched on both sides of the shallow-etched Si waveguide rib with a 170 nm gap separation to the Si waveguide core. 
The fully etched grating provides the optical resonant feedback for the pump laser and determines the lasing wavelength \(\lambda_{\text {pump}}\) by its pitch \(\Lambda\) (\(\lambda_{\text {pump}}=2 n_{\text {eff}} \Lambda\)). 
Here $\Lambda=238$ nm and $\lambda_{\text {pump}}\sim1550$ nm. 
A quarter wavelength shift section is included at the grating length center to supply non-degenerate phase conditions favoring single longitudinal mode lasing. 

The single-wavelength laser output passes through a thermo-optic resistive heater (for optical phase control), and couples into a high-$Q$ Si$_3$N$_4$ microring resonator where nonlinear four-wave mixing processes generate Kerr microcombs\cite{Kippenberg:04, DelHaye:07}. 
The Si$_3$N$_4$ microresonator exhibits anomalous group velocity dispersion (GVD) in the telecommunication C band and have a free-spectral range (FSR) of 100 GHz.
The laser directly pumps the microresonator without an intermediate optical isolator, and the entire device is operated via laser control and phase control.
Laser self-injection locking \cite{Kondratiev:17, Liang:15b, Liang:15, Raja:19, Shen:20, jin2021hertz} leverages the narrow-band optical feedback at desired phase relations from a high-$Q$ microresonator to stabilize the pump laser and pulls the laser frequency towards the microring resonance. 
In this scenario, soliton microcombs can form when optimum laser-resonator frequency detuning is reached. 
Only the laser and phase tuner currents are electronically controlled, and no sophisticated external electronics feedback controls are required. 
The DFB laser wavelength increases with increasing laser current, as the grating index increases due to injected electrical power heating. 
As a result, certain gain currents triggers comb generation if the laser wavelength coincides with a microresonator resonance. 
The comb generation region resides where the laser is red-detuned to the resonance and the phase of the back-scattered light from the microresonator to the laser fulfills certain phase relations (Fig. \ref{Fig:1}E, see Supplementary Information for details).
Note that the back-scattered light originates from Rayleigh scattering inside the microresonator due to surface roughness and bulk inhomogeneity. 

\textbf{Fabrication process flow}. 
Figure \ref{Fig:2}A presents the wafer-scale fabrication process flow. 
It starts with the photonic Damascene process \cite{Liu:20b} to fabricate the Si$_3$N$_4$ PIC on a Si substrate with 4-{\textmu}m-thick thermal wet SiO$_2$. 
The Si$_3$N$_4$ PIC is exposed with DUV stepper lithography and transferred to the SiO$_2$ substrate to form the waveguide preform. 
Stoichiometric Si$_3$N$_4$ is deposited on the patterned SiO$_2$ preform using low-pressure chemical vapor deposition (LPCVD), filling the trenches and forming waveguide cores. 
Chemical-mechanical polishing (CMP) is used to remove excess Si$_3$N$_4$ and planarize the wafer front surface, followed by spacer SiO$_2$ deposition of 300 nm thickness on the Si$_3$N$_4$ substrate. 
The entire substrate is further annealed at 1200$^\circ$C to drive out the residual hydrogen content in the Si$_3$N$_4$ and SiO$_2$ films, and to densify the spacer SiO$_2$. 

A second CMP is performed to create a flat and smooth wafer surface. 
As shown in Fig. \ref{Fig:2}B left, the measured RMS roughness of the wafer surface using atomic-force microscopy (AFM) is 0.27 nm, enabling direct substrate bonding with an SOI wafer. 
After removing the Si substrate of the bonded SOI wafer, the Si device layer is processed to create waveguide structures with different etch depths, including shallow-etched Si waveguides for the lasers and phase tuners, fully-etched hole structures for gratings (cf. Fig. \ref{Fig:2}B right), and thin-thickness Si tapers for mode conversion between the Si waveguide and underlying Si$_3$N$_4$ waveguide\cite{Xiang:20}. 
InP MQW epi is then bonded to the patterned Si device surface at the active regions. 
The InP process starts with InP substrate removal. 
InP mesa etches including P-type InP, AlInGaAs MQW, and N-type InP etches are performed using selective dry etching and wet etching. 
The P- and N-type contact metals are deposited on the P-InGaAs layer and N-InP layer respectively. 
The excess Si on top of the Si$_3$N$_4$ microresonators is removed before laser passivation using hydrogen-free deuterated SiO$_2$ deposition\cite{Jin:20}. 
Vias are then etched through it for laser electrical contact. 
Proton implantation is performed on the laser mesa structure to reduce electrical current leakage. 
Heater and probe metals are deposited at the end of the full process.
Finally, the entire wafer is diced into dozens of dies/chips to facilitate testing. 
Each chip contains tens of devices for soliton generation. 
Figure \ref{Fig:2}C shows the scanning electron microscopic (SEM) images of the device cross-section, which is false-colored to illustrate the multilayer structure. 

%%%%%%%%%%%%%%%%%%%%%%%%%%%%%%%%%%%%%%%%%%%%%%%%%%%%%%%%%%%%%%%%%%%%%%
\begin{figure*}[t!]
\centering
\includegraphics{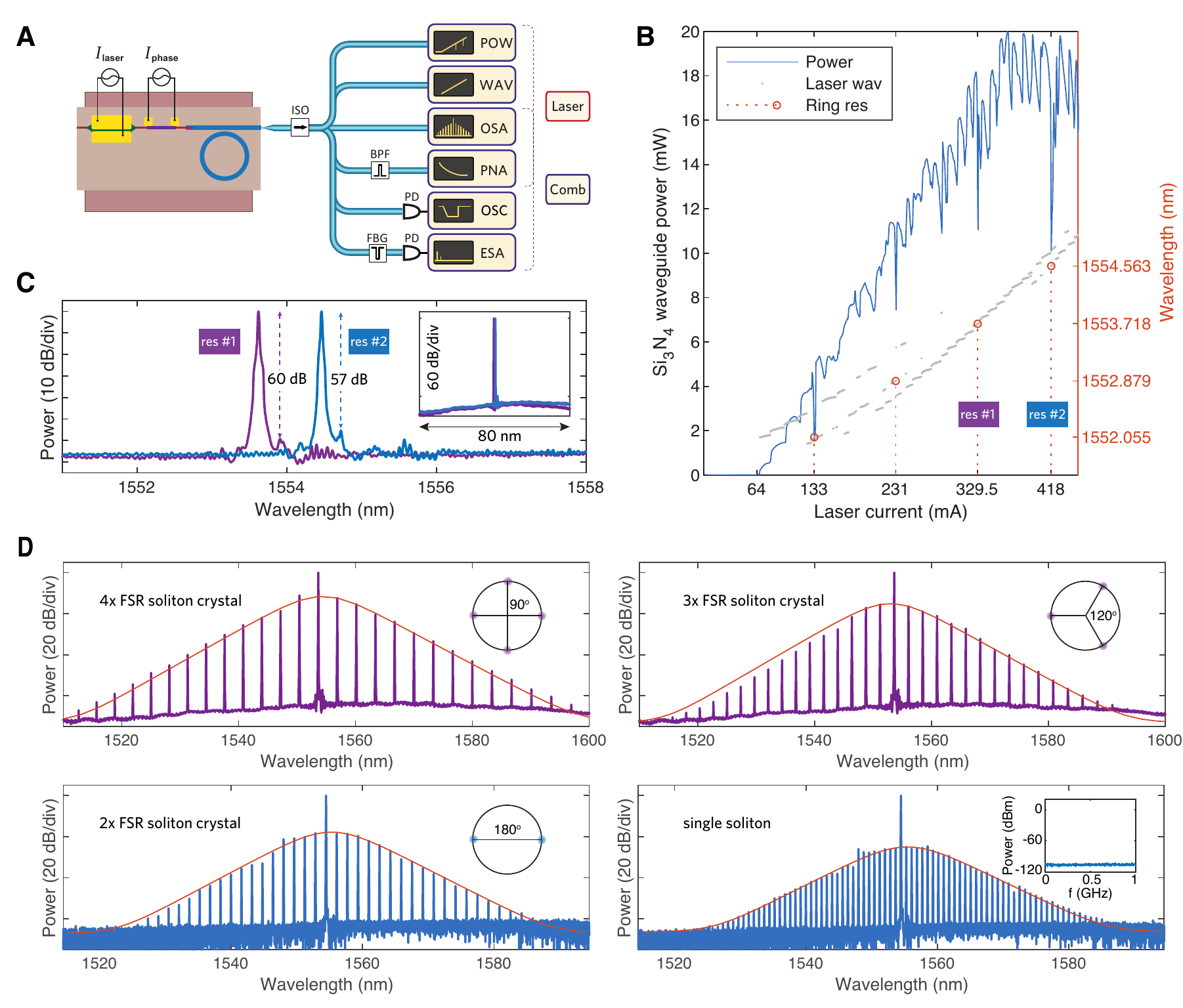}
\caption{\textbf{Device characterization and experimental generation of soliton spectra}.
A. Schematics of the experimental setup for laser and soliton characterization. 
$I_\text{laser}$ and $I_\text{phase}$ are the current sources to drive the laser and the phase tuner. 
POW: power meter. 
WAV: wavelength meter. 
OSA: optical spectrum analyzer. 
PNA: phase noise analyzer.  
OSC: oscilloscope.  
ESA: electrical spectrum analyzer.  
ISO: isolator. 
BPF: band-pass filter. 
PD: photodetector. 
B. Light-current sweep measurement with stepped laser current and fixed phase tuner current. 
Grey color shows the corresponding laser center wavelength as a function of the laser current. 
Red circles indicate the laser wavelength coinciding with microresonator resonances. 
Soliton microcombs are generated with laser currents and wavelengths at resonance \#1 and resonance \#2.  
C. Single-mode DFB laser spectra at the wavelengths of resonance \#1 and resonance \#2.  
D. Optical spectra of soliton microcombs. 
Inset shows the relative position of multi-solitons circulating inside the microring resonator for 4-, 3-, and 2-FSR soliton crystal states and low-frequency RF spectrum of the single soliton state.
}
\label{Fig:3}
\end{figure*}
%%%%%%%%%%%%%%%%%%%%%%%%%%%%%%%%%%%%%%%%%%%%%%%%%%%%%%%%%%%%%%%%%%%%%%

%%%%%%%%%%%%%%%%%%%%%%%%%%%%%%%%%%%%%%%%%%%%%%%%%%%%%%%%%%%%%%%%%%%%%%
\begin{figure*}[t!]
\centering
\includegraphics{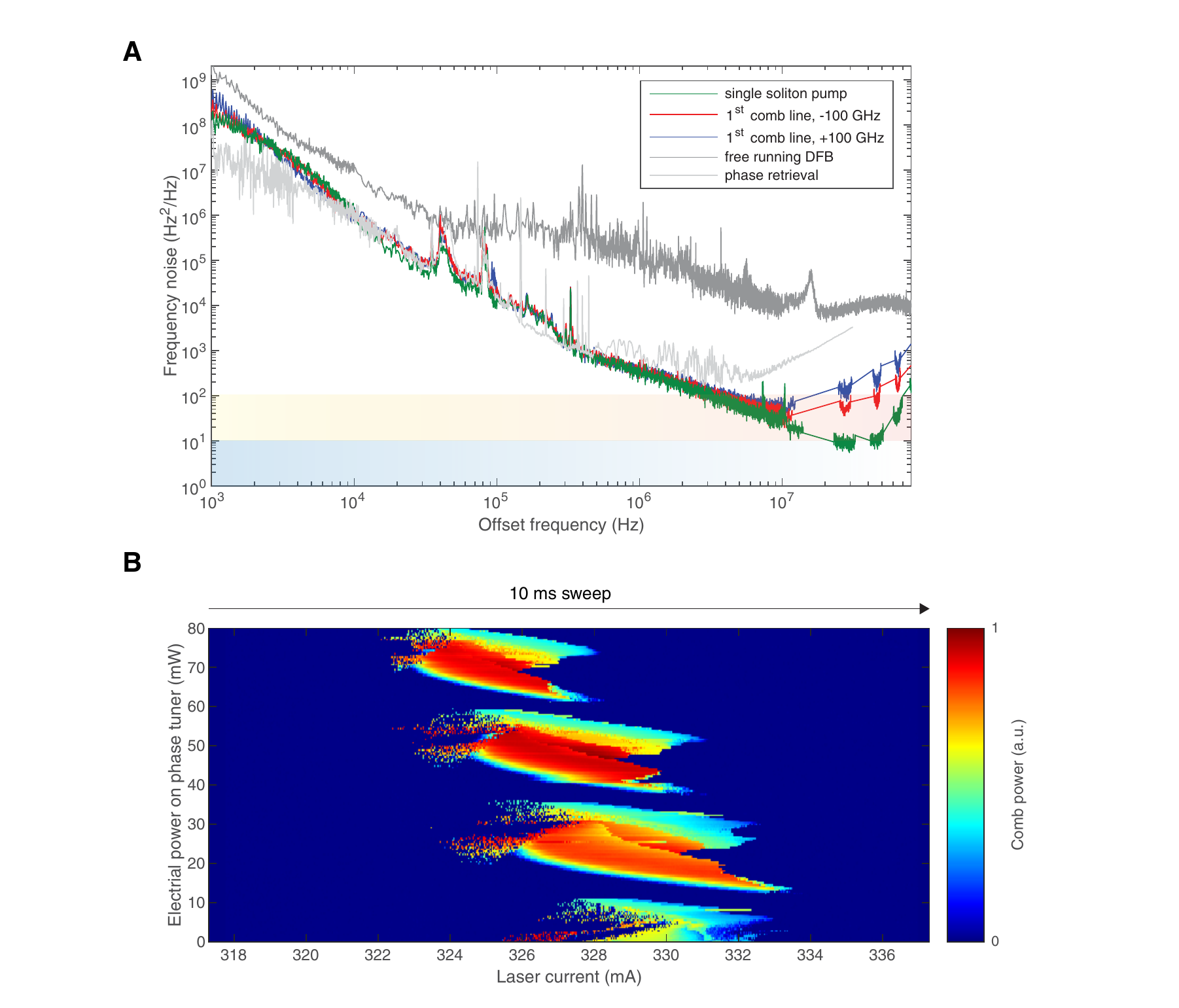}
\caption{\textbf{Laser frequency noise spectrum and comb generation with self-injection locking}.
A. Frequency noise spectra of the self-injection-locked pump line (green), comb lines with $\pm 100$ GHz frequency offset to the pump (blue and red) in the single soliton state, free-running single-mode DFB laser output without self-injection locking (dark grey), and comb line around 1550 nm for the 2-FSR soliton crystal state using optical phase retrieval method (see Supplementary Information). 
B. Comb power evolution with sweeping laser current (10 milliseconds sweep) under varying electrical power on the phase tuner.
}
\label{Fig:4}
\end{figure*}
%%%%%%%%%%%%%%%%%%%%%%%%%%%%%%%%%%%%%%%%%%%%%%%%%%%%%%%%%%%%%%%%%%%%%%

\textbf{Soliton generation}.
The experimental setup to characterize the final chip devices is shown in Fig. \ref{Fig:3}A. 
We first characterize the laser performance. 
Figure \ref{Fig:3}B shows the measured light-current curves (solid blue curve), i.e.  the laser output power versus stepped laser injection current of 0.5 mA, fixed phase tuner current of 7 mA, and 20$^\circ$C stage temperature. 
The measured laser threshold current is 64 mA. 
The laser power is out-coupled from the Si$_3$N$_4$ waveguide inverse-taper to a lensed fiber. 
The maximum power in the output fiber is measured as 8 mW, and the corresponding on-chip power in the Si$_3$N$_4$ waveguide is approximately 20 mW. 
The measured laser center wavelength using a wavelength meter is shown in Fig. \ref{Fig:3}B (dashed grey curve). 
Several dips on the optical power are observed at laser currents of 133, 231, 329.5, and 418 mA, where the laser wavelength coincides with microresonator resonances. 
These dips also verify the microresonator FSR of 100 GHz. 
Note that a high-reflection (HR) coating is applied on the other side of the DFB laser to boost the laser output power. 
Future devices can avoid using this coating to yield mode-hop-free DFB lasers and linearized laser wavelength dependence on the laser current. 
The DFB laser has high SMSR at resonance \#1 and \#2, where high output powers are also obtained that are advantageous for comb generation. 
Figure \ref{Fig:3}C shows the single-longitudinal-mode laser spectra at the two resonance wavelengths, with 60 and 57 dB SMSR, respectively.

Soliton microcomb is generated by tuning the laser frequency to the microresonator resonance via biasing the laser current, together with tuning the current of the phase tuner to control the relative forward/backward phase relations. 
As no optical isolator is used between the laser and the Si$_3$N$_4$ microresonator, laser self-injection locking\cite{Kondratiev:17, Liang:15b, Shen:20} occurs when the laser frequency coincides with a microresonator resonance.
More information of the phase dependence on comb generation is revealed by simulating the nonlinear self-injection locking process\cite{Voloshin:21} and is discussed in the Supplementary Information. 
The calculated Kerr parametric oscillation threshold \cite{Kippenberg:04} for Si$_3$N$_4$ microresonators of $Q_0\sim7\times10^6$ (see Supplementary information for $Q$ measurement) and 100 GHz FSR, is estimated to be around 3 mW. 
With around 329 mA laser current, the estimated on-chip laser power to pump the Si$_3$N$_4$ microresonator exceeds 16 mW. 
As shown in Fig. \ref{Fig:3}D, perfect soliton crystal states\cite{Karpov:19} are observed by fine-tuning the laser frequency. 
We also observe soliton crystal states with decreasing crystallization orders when increasing the laser current (red-detuning). 
Further increasing the laser power to around 418 mA current, a two-soliton crystal state with 200 GHz repetition rate and a single-soliton state with 100 GHz repetition rate is generated. 
The coherent soliton nature is revealed by photodetection of the amplitude noise of the comb lines beat signal. 
Once generated, the soliton states can be stable for hours in standard lab environments without any external feedback control. 
This stability benefits from the monolithic nature of the chip device and the laser-microresonator coupling through laser self-injection locking. 

\textbf{Self-injection locking and phase noise reduction}.
To further characterize the laser self-injection locking, we measured the frequency noise spectra of the pump line and comb line in the single soliton state and laser free-running state, as shown in Fig. \ref{Fig:4}A.
Leveraging a high-$Q$ external cavity, self-injection locking exhibits frequency noise reduction, i.e. laser linewidth narrowing \cite{Liang:15b}. 
The fundamental linewidth of the free-running DFB laser is around 60 kHz, and is reduced to about 25 Hz in the self-injection-locked single soliton state. 
A 10-dB noise reduction is observed at 1 kHz Fourier offset frequency and is further increased to more than 30 dB above 300 kHz offset frequency. 
It has been revealed recently that the frequency noise of a self-injection-locked laser to a high-$Q$ Si$_3$N$_4$ microresonator can be dominantly limited by the thermo-refractive noise of the microresonator \cite{Huang:19, jin2021hertz}.
Therefore, further noise reduction can be realized by using microresonators of larger sizes (i.e. smaller FSR) or engineered material properties with lower thermo-optic transduction.
Though the free-running linewidth of the DFB laser is broad, due to the linewidth narrowing provided by self-injection locking, the laser is still able to directly generate soliton states, surpassing its intrinsic limitation of spectral impurity. 
Additionally, the coherence of the injection-locked pump line is transferred to other comb lines. 
For example, the first neighboring comb lines, 100 GHz apart from the pump, have a fundamental linewidth around 200 to 300 Hz, and their frequency noise below 10 MHz offset frequency directly inherits that of the self-injection-locked pump. 
The self-injection locking scheme thus enables multi-wavelength, narrow-linewidth laser sources. 
Our device represents the first demonstration of a self-injection-locked, narrow-linewidth laser on Si.

Different from the conventional pumping scheme with an optical isolator, the comb generation in the self-injection locking scheme relies critically on the phase relation between the forward signal (i.e. the laser emission to the microresonator) and backward signal (i.e. the back-scattered light from the microresonator to the laser). 
In this case, comb generation is only allowed when the optical phase difference is within a certain optimized range. 
In previous works using hybrid integrated devices, controlling the phase difference is realized by varying the gap distance between the laser chip and Si$_3$N$_4$ chip, which however also causes power coupling efficiency fluctuation between these two chips. 
In our monolithic device, this optical phase difference can be directly controlled by varying the current to the phase tuner. 
We experimentally studied the comb formation with laser current sweep under different currents on the phase tuner. 
In order to exclude the interference from the mode-hop phenomenon when monitoring the power of the new frequency components, we sweep the laser current across the resonance \#1 shown in Fig. \ref{Fig:3}B. 
Results shown in Fig. \ref{Fig:4}B unveils that the comb generation is only permitted with certain phase conditions. 
More study shown in the Supplementary Information indicates that the back-scattered signal needs to be in phase with the forward signal. 
Additionally, the phase altering effect is periodic and deterministic with the applied electrical power on the phase tuner. 
Thus the allowed comb generation area depends mainly on the pump power and the intensity of the back-scattered light (see Supplementary Information). 

In summary, we have presented the first heterogeneously integrated laser soliton microcomb on silicon. 
Thousands of devices are manufactured from a full wafer-scale fabrication process using standard CMOS techniques such as DUV stepper lithography, CMP, wafer bonding, etc. 
We have successfully demonstrated single soliton formation with 100 GHz repetition rate, and characterized laser performance with self-injection locking. 
More functions could be added based on our current technology. 
For example, active, high-speed tuning of the Si$_3$N$_4$ microresonators (thus the injection-locked laser frequency) can be realized using piezoelectric actuators \cite{Jin:18, Liu:20a}. 
Meanwhile, full integration with other nonlinear photonics platforms such as (Al)GaAs \cite{Chang:20,chang_strong_2019} and LiNbO$_3$ \cite{Zhang:19,boes_status_2018,HeM:19} could provide $\chi(2)$ nonlinearity that can be used for electro-optic modulation and second harmonic generation. 
The laser gain material can also be modified to utilize quantum-dots \cite{Liu:15} or extended to different wavelength ranges from the visible to mid-infrared, significantly complementing existing integrated nonlinear photonic systems.
Though demonstrated here on 100-mm-diameter Si substrates, our process could be upgraded to 200- or 300-mm-diameter substrates using modified CMOS foundry pilot lines. 
Our work paves the way for large-volume, low-cost manufacturing of electrically controlled, chip-based frequency comb modules for future applications in data-centers, space and mobile platforms.

\smallskip
\begin{footnotesize}

\noindent \textbf{Acknowledgments}: 
We thank Mario Dumont and Jijun He for their help in the device characterization.
This work is supported by the Defense Advanced Research Projects Agency (DARPA) under DODOS (HR0011-15-C-055) programmes of the Microsystems Technology Office (MTO).

\smallskip

\noindent \textbf{Author contribution}:
C.X., L.C., J.L., T.J.K. and J.E.B. conceived the idea and initiated the collaboration.
C.X., J.L. and L.C. designed the wafer layout and process flow. 
J.L. and R.N.W. fabricated the Si$_3$N$_4$ substrate. 
C.X. and J.P. fabricated the heterogeneous InP/Si lasers and Si circuits with the assistance from W.X., and Z.Z. 
C.X., J.G., and J.L. tested the final chips. 
C.X. and J.G. performed the laser soliton experiments with the assistance from J.R..
W.W. performed the numerical simulations of self-injection locking.
J.S. took the FIB-SEM image.
All authors discussed the data.
C.X., J.L. and W.W. wrote the manuscript, with input from others. 
T.J.K. and J.E.B supervised the project.

\smallskip

\noindent \textbf{Data Availability Statement}: 
All data generated or analysed during this study are available within the paper and its Supplementary Information. 
Further source data will be made available on reasonable request.

\end{footnotesize}

\bibliographystyle{apsrev4-1}
\bibliography{bibliography}
\end{document}

\title{Supplementary Information to:  Laser soliton microcombs on silicon}

\author{Chao Xiang}
\thanks{These authors contributed equally to this work.}
\affiliation{ECE Department, University of California Santa Barbara, Santa Barbara, CA 93106, USA}

\author{Junqiu Liu}
\thanks{These authors contributed equally to this work.}
\affiliation{Institute of Physics, Swiss Federal Institute of Technology Lausanne (EPFL), CH-1015 Lausanne, Switzerland}

\author{Joel Guo}
\affiliation{ECE Department, University of California Santa Barbara, Santa Barbara, CA 93106, USA}

\author{Lin Chang}
\affiliation{ECE Department, University of California Santa Barbara, Santa Barbara, CA 93106, USA}

\author{Rui Ning Wang}
\affiliation{Institute of Physics, Swiss Federal Institute of Technology Lausanne (EPFL), CH-1015 Lausanne, Switzerland}

\author{Wenle Weng}
\affiliation{Institute of Physics, Swiss Federal Institute of Technology Lausanne (EPFL), CH-1015 Lausanne, Switzerland}

\author{Jonathan Peters}
\affiliation{ECE Department, University of California Santa Barbara, Santa Barbara, CA 93106, USA}

\author{Weiqiang Xie}
\affiliation{ECE Department, University of California Santa Barbara, Santa Barbara, CA 93106, USA}

\author{Zeyu Zhang}
\affiliation{ECE Department, University of California Santa Barbara, Santa Barbara, CA 93106, USA}

\author{Johann Riemensberger}
\affiliation{Institute of Physics, Swiss Federal Institute of Technology Lausanne (EPFL), CH-1015 Lausanne, Switzerland}

\author{Jennifer Selvidge}
\affiliation{Materials Department, University of California Santa Barbara, Santa Barbara, CA 93106, USA}

\author{Tobias J. Kippenberg}
\email[]{tobias.kippenberg@epfl.ch}
\affiliation{Institute of Physics, Swiss Federal Institute of Technology Lausanne (EPFL), CH-1015 Lausanne, Switzerland}

\author{John E. Bowers}
\email[]{bowers@ece.ucsb.edu}
\affiliation{ECE Department, University of California Santa Barbara, Santa Barbara, CA 93106, USA}
\affiliation{Materials Department, University of California Santa Barbara, Santa Barbara, CA 93106, USA}

\maketitle
\section{Detailed Device Fabrication}

\setcounter{figure}{0} 

%%%%%%%%%%%%%%%%%%%%%%%%%%%%%%%%%%
\begin{figure*}[b!]
\renewcommand{\figurename}{\textbf{Supplementary Figure}}
\centering
\includegraphics{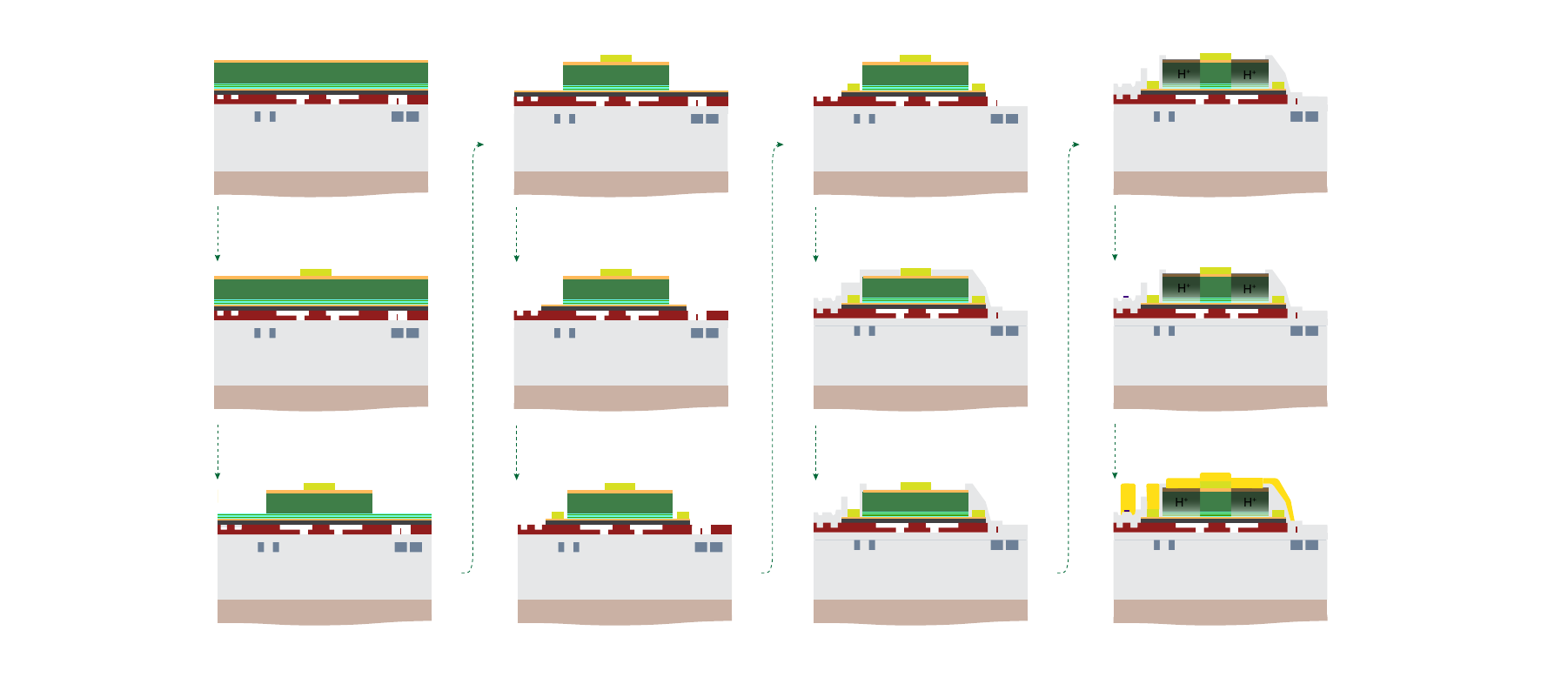}

\caption{
\textbf{Extended InP processing flow}. 
From the left to right columns: 
InP bonding and substrate removal (top), P-contact metal deposition (middle),  P-type InP etch (bottom);  
MQW etch (top), N-type InP etch (middle), N-contact metal deposition (bottom); 
Excess Si removal (top), via oxide deposition (middle), via open etch (bottom); 
Proton implantation (top), heater metal deposition (middle), probe metal deposition (bottom).}
\label{Fig:SI01}
\end{figure*}
%%%%%%%%%%%%%%%%%%%%%%%%%%%%%%%%%%

The full fabrication process of laser soliton microcomb devices starts with a 100 mm-diameter, patterned Si$_3$N$_4$ wafer fabricated using the photonic Damascene process, as described by the main text. 
After the second CMP on the SiO$_2$ spacer, the wafer front-side surface is sufficiently smooth with less than 0.3 nm RMS roughness. 
Vertical channels for outgassing, comprising bars of 1 $\mu$m width, are etched with 500 nm depth in the open zoom on the SiO$_2$\cite{liang_highly_2008} spacer. 
After RCA cleaning, we bond a $60\times60$ mm$^2$ SOI piece on the Si$_3$N$_4$ wafer using plasma-assisted direct wafer bonding. 
Larger SOI pieces can be used in the future to enable more device throughput. 
After 2 hours annealing of the bonded substrate at 300$^\circ$C, the Si substrate of the SOI piece is mechanically lapped down to around 80 $\mu$m thickness, followed by Si Bosch process to remove the remaining Si substrate. 
The 1 $\mu$m thick buried oxide layer is then removed by buffered hydrofluoric acid, leaving only the Si device layer on the Si$_3$N$_4$ substrate for the following steps.

The Si device layer is 500 nm thick. 
DUV-stepper lithography (ASML PAS 5500/300, 248 nm) is used to pattern the Si devices, which are then etched into the Si layer from the photoresist mask. 
Three etch depths are used for the Si devices. 
The hybrid InP/Si section, phase tuner section, and Si waveguide section have an etch depth around 231 nm, which is controlled during the Si etch using an etch monitor (Intellemetrics LEP50). 
The 231 nm etch depth thus creates a 269-nm-thick Si layer. 
On this layer, the Si waveguide is etched and tapered from 400 nm to 200 nm wide for the mode transition between the Si waveguide and the Si$_3$N$_4$ waveguide underneath. 
Electron beam lithography (EBL) is used to pattern the grating holes on this layer alongside the hybrid InP/Si waveguide section, with 238 nm pitch size and a fill factor around 0.5. 
The final step for the Si layer processing is to etch 500 nm deep vertical outgassing channels for bonding with InP MQW epi wafers. 

Multiple cleaved InP MQW epi dies are selectively bonded to active regions covering the laser gain areas. 
The bonded wafer is then annealed at 300$^\circ$C again for 2 hours, and InP substrate material is mechanically lapped down to 60 $\mu$m. 
The remaining InP substrate removal is performed by using 3:1 hydrochloric acid (HCl)/H$_2$O, such that the wet-etch stops at the P-doped InGaAs layer. 
InP processing then starts with P-type contact metal deposition (the layer stack is Pd/Ti/Pd/Au with 3/17/17/200 nm thickness, respectively). 
Afterwards, the InP mesa undergoes three etches: P-type InP dry etch using methane/hydrogen/argon, MQW wet etch using H$_2$O/ hydrogen peroxide (H$_2$O$_2$)/ phosphoric acid (H$_3$PO$_4$) solution, and then N-type InP dry etch using methane/hydrogen/argon. 
After the InP mesa formation, N-type contact metal (Pd/Ge/Pd/Pt/Au, of 10/110/25/200/1000 nm thickness, respectively) and a second-layer P metal (Ti/Pt/Au, 20/200/1300 nm) are deposited. 
In addition to Si/Si$_3$N$_4$ tapers, excess bonded Si on top of Si$_3$N$_4$ is removed by XeF$_2$ isotropic gas etch. 
Deuterated SiO$_2$ of around 900 nm thickness is then deposited \cite{Jin:20} as the via oxide, which is opened at selected P- and N-metal regions by SiO$_2$ dry etch. 
Proton implantation follows the via oxide open at the InP mesa region. 
Heaters (Ti/Pt, 10/100 nm) and probe metals (Ti/Au, 23/1500 nm) are deposited. 
Finally, the entire 4-inch wafer is diced into multiples dies for testing, followed by the laser facet polishing. 
During the process, all the metal depositions are performed using electron beam deposition. 
The extended InP processing with cross-sectional views is shown in Supplementary Fig. \ref{Fig:SI01}.

%%%%%%%%%%%%%%%%%%%%%%%%%%%%%%%%%%%%%%%%%%%%%%%%%%%%%%%%%%%%%%%%%%%%%%%%%%%%%%%%%%%%%%%%%%%%%%%%%%%%%%
\section{Measurements of Si$_3$N$_4$ microresonator $Q$}

%%%%%%%%%%%%%%%%%%%%%%%%%%%%%%%%%%
\begin{figure*}[b!]
\renewcommand{\figurename}{\textbf{Supplementary Figure}}
\centering
\includegraphics[width=1\textwidth]{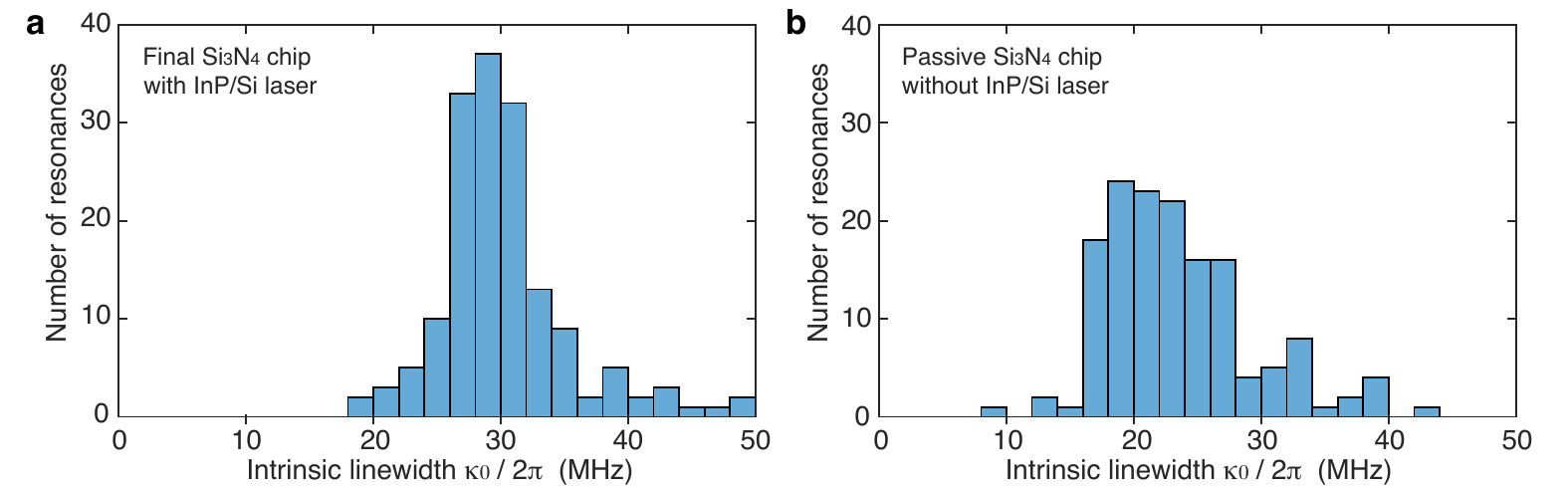} 
\caption{
\textbf{Histogram of measured intrinsic loss $\kappa_0/2\pi$ of Si$_3$N$_4$ microresonators at different stage during fabrication.}
\textbf{(a)} For the Si$_3$N$_4$ chip with integrated InP/Si lasers, the most probable value of $\kappa_0/2\pi$ is 29 MHz, corresponding to an estimated $Q_0\approx6.8\times10^6$. 
\textbf{(b)} For the Si$_3$N$_4$ chip without integrated InP/Si lasers (i.e. purely passive devices), the most probable value of $\kappa_0/2\pi$ is 19 MHz, corresponding to an estimated $Q_0\approx10.2\times10^6$. 
Therefore, $Q$ degradation due to the back-end InP/Si laser fabrication is observed and with further optimization the soliton threshold can be lowered in future devices.}
\label{figS3}
\end{figure*}
%%%%%%%%%%%%%%%%%%%%%%%%%%%%%%%%%%

The threshold power for Kerr parametric oscillation \cite{Kippenberg:04} is proportional to $1/Q^2$. 
Therefore for soliton microcomb generation, it is critical to achieve high microresonator $Q$. 
This requires: 1. Using the optimized photonic Damascene process \cite{Pfeiffer:18b, Liu:20b} to achieve the lowest loss possible (thus the highest $Q$) in the Si$_3$N$_4$ PIC before the SOI bonding; and 2. Minimizing extra loss introduced during the back-end InP/Si laser processing. 
Both require us to carefully design and perform the fabrication process. 

To measure the Si$_3$N$_4$ microresonator $Q$ factor of the final chip devices with integrated InP/Si lasers, chips are diced such that light can be coupled into and out of the Si$_3$N$_4$ bus waveguide via edge coupling with input and output lensed fibers. 
We use the frequency-comb-assisted diode laser spectroscopy \cite{DelHaye:09, Liu:16} to characterize the optical resonances of the Si$_3$N$_4$ microresonators and calculate the $Q$ factors. 
The optical transmission spectrum of the microresonator is calibrated with a self-referenced, fiber-laser-based optical frequency comb. 
Optical resonances are identified and fitted \cite{Liu:18a}, enabling extraction of the precise frequency of each resonance $\omega/2\pi$, as well as the resonance's loaded (full) linewidth $\kappa/2\pi=\kappa_0/2\pi+\kappa_\text{ex}/2\pi$, the intrinsic loss $\kappa_0/2\pi$ and the bus-waveguide-to-microresonator external coupling strength $\kappa_\text{ex}/2\pi$ \cite{Pfeiffer:17}. 
Here we mainly analyze the intrinsic loss $\kappa_0/2\pi$ that directly reflects the quality of our fabrication process. 
As shown in Supplementary Fig. \ref{figS3}(a), for the fundamental transverse-electric mode (TE$_{00}$ mode) of the microresonator, the histograms of $\kappa_0/2\pi$ presents a most probable value of 29 MHz, corresponding to an estimated $Q_0\approx6.8\times10^6$. 
To understand how much the $Q$ degrades due to the back-end InP/Si laser fabrication, we also separate a purely passive Si$_3$N$_4$ substrate whose fabrication is completed before the SOI bonding. 
For chips from this substrate, as shown in Supplementary Fig. \ref{figS3}(b), the histograms of measured $\kappa_0/2\pi$ presents a most probable value of 19.5 MHz, corresponding to an estimated $Q_0\approx10.2\times10^6$. 
We note that, due to the absence of top SiO$_2$ cladding in the current case, the $Q_0$ value is lower than our previous results (e.g. Ref. \cite{Liu:20b}, $Q_0>30\times10^6$) using fully SiO$_2$-cladded Si$_3$N$_4$. 
Therefore, with this comparison of $\kappa_0/2\pi$ values, we conclude that the microresonator $Q$ indeed degrades due to the following InP/Si laser process. 
Nevertheless, the current $Q$ value is sufficient for soliton generation of 100 GHz repetition rate. 
For future work requiring higher $Q$ for e.g. microwave-repetition-rate solitons \cite{Liu:20}, thicker SiO$_2$ spacer between the Si$_3$N$_4$ layer and Si layer is desired. 
In addition, Si removal and cleaning methods can be further optimized. 

%%%%%%%%%%%%%%%%%%%%%%%%%%%%%%%%%%%%%%%%%%%%%%%%%%%%%%%%%%%%%%%%%%%%%%%%%%%%%%%%%%%%%%%%%%%%%%%%%%%%%%
\section{Phase noise measurement using phase retrieval method}

The laser phase noise spectra of the center laser line (pump) and the comb line are measured with a commercial phase noise analysis system (PNA, OEWaves4000) and plotted in Fig. 4 of the main manuscript. 
We find that the suppression of phase and frequency noises due to the self-injection locking depends strongly on the offset frequency, and ranges from 10~dB at 1~kHz Fourier offset frequency to 30~dB at offset frequencies above 10~MHz. 
The PNA system uses the delayed-homodyne technique to determine the laser phase and frequency noises. 
It is well known that such systems have limitations in the measurement of narrow-band laser linewidths, and introduce additional high-frequency peaks that are harmonics of the free spectral range of the employed interferometer\cite{ludvigsen1998laser}. 
In order to benchmark and verify the results of the PNA, we perform a second measurement of the self-injection-locked laser frequency noise by recording a heterodyne beatnote with a second low-noise laser (OEWaves OE4030) on a fast photodetector. 
The 41~GHz beatnote is down-mixed to 60~MHz and sampled with a low-noise sampling oscilloscope at 125~MHz. A high-pass filter removes the laser intensity noises. 
The quadature signal of this beatnote is reconstructed via the Hilbert transform
\begin{equation}
\Phi(t) = \arctan{\dfrac{H(U(t))}{U(t)}}
\end{equation}
The instantaneous phase can be inferred from the In-phase and Quadrature components directly, and the carrier frequency of the heterodyne beat is removed by subtraction of a linear function. 
The power spectral density of the phase noise is obtained using Welch's method of windowed Fourier transform and successive averaging. 

%%%%%%%%%%%%%%%%%%%%%%%%%%%%%%%%%%%%%%%%%%%%%%%%%%%%%%%%%%%%%%%%%%%%%%%%%%%%%%%%%%%%%%%%%%%%%%%%%%%%%%
\section{Simulation of microcomb generation via laser self-injection locking}

%%%%%%%%%%%%%%%%%%%%%%%%%%%%%%%%%%
\begin{table}[b!]
\centering
\begin{tabular}{|c|c|c|c|}
\hline
Symbol & Value & Unit & Definition \\
\hline
$\alpha$ & $5$   &                 & Linewidth enhancement factor \\
$a$ & $1\times10^{4}$   & s$^{-1}$       & Differential gain \\
$N_0$ & $1\times10^{24}$   & m$^{-3}$                & Carrier density at transparency \\
$\kappa$ & $1\times10^{11}\times2\pi$   & rad$\cdot$s$^{-1}$         & Laser cavity loss rate\\
$I$ & $0.24$   & A                & Laser biased current \\
$\gamma$ & $1\times10^9$         & s$^{-1}$           & Carrier recombination rate\\
$V$ & $2\times10^{-16}$   & m$^{3}$                & Volume of active section \\
$e$ & $1.6\times10^{-19}$   & C                & Elementary electronic charge \\
%\hline
$\kappa_\mathrm{r}$ & $150\times10^6\times2\pi$   &  rad$\cdot$s$^{-1}$    & Loaded loss rate \\
$\kappa_\mathrm{sc}$ & $20\times10^6\times2\pi$  &  rad$\cdot$s$^{-1}$  &cw-ccw coupling rate \\
$\kappa_\mathrm{inj}$ & $4.75\times10^{8}\times2\pi$   &  rad$\cdot$s$^{-1}$  & Laser-microresonator coupling rate \\
$g$ &    $1.866$         & rad$^{-1}$     & single-photon induced Kerr shift \\
$D_\mathrm{2}$ & $1\times10^{6}\times2\pi$         &  rad$^2\cdot$s$^{-1}$      & Dispersion coefficient \\
$\eta$ & $2.676\times10^{-7}$         &    m$^{\frac{3}{2}}$    & conversion factor \\
\hline
\end{tabular}
\caption{Values and definitions of parameters used in the simulations.}
  \label{table1}
\end{table}
%%%%%%%%%%%%%%%%%%%%%%%%%%%%%%%%%%

We numerically simulate the dynamics of the self-injection-locked-laser-microcomb system using coupled equations that include both the semiconductor laser rate equations and the coupled intracavity field equations for counter-circulating fields in the Kerr-nonlinear microresonator. 
The coupled equations are written as:
%%%%%%%%%%%%%%%%%%%%%%%%%%%%%%
\begin{equation}
\label{eq1}
\frac{dN}{dt} = \frac{I}{eV} - \gamma N - aV(N-N_0) |E_\mathrm{laser}|^2
\end{equation}
%%%%%%%%%%%%%%%%%%%%%%%%%%%%%%
\begin{equation}
\label{eq2}
\frac{dE_\mathrm{laser}}{dt} = \left[\frac{1}{2} (1 - i \alpha) (a V (N-N_0) - \kappa) -i\Delta\omega \right]E_\mathrm{laser}+ \eta^{-1} \kappa_\mathrm{inj} e^{i \theta} A_\mathrm{ccw}
\end{equation}
%%%%%%%%%%%%%%%%%%%%%%%%%%%%%%
\begin{equation}
\label{eq3}
\frac{\partial{A_\mathrm{cw}}}{\partial t} - i \frac{1}{2} D_{2} \frac{\partial^2{A_\mathrm{cw}}}{\partial \phi^2} - i g (|A_\mathrm{cw}|^2 + 2 |A_\mathrm{ccw}|^2) A_\mathrm{cw}  = \\
-\left( {\frac{\kappa_\mathrm{r}}{2} + i(\omega_0 - \omega_\mathrm{laser}) } \right){A_\mathrm{cw}} \\
+ i \kappa_\mathrm{sc} A_\mathrm{ccw} + \eta \kappa_\mathrm{inj} e^{i \theta} E_\mathrm{laser}
\end{equation}
%%%%%%%%%%%%%%%%%%%%%%%%%%%%%%
\begin{equation}
\label{eq4}
\frac{dA_\mathrm{ccw}}{dt} - i g \left(|A_\mathrm{ccw}|^2 + 2 \int_{0}^{2\pi}\frac{|A_\mathrm{cw}|^2}{2\pi} \,d\phi\right) A_\mathrm{ccw} = \\
-\left( {\frac{\kappa_\mathrm{r}}{2} + i(\omega_0 - \omega_\mathrm{laser}) } \right){A_\mathrm{ccw}} \\
+ i \kappa_\mathrm{sc}^* A_\mathrm{cw}^0
\end{equation}
%%%%%%%%%%%%%%%%%%%%%%%%%%%%%%
where $E_\mathrm{laser}$ describes the complex laser field, $N$ is the carrier density, $\alpha$ is the linewidth enhancement factor, $a$ is the differential gain, $V$ is the laser active volume, $N_0$ is the carrier density at transparency, $\gamma$ is the carrier recombination rate, and $e$ is the elementary charge. 
For the microresonator, $A_\mathrm{cw}$ and $A_\mathrm{ccw}$ are the slowly varying field amplitudes of the clockwise (cw) and the counterclockwise (ccw) modes respectively, $\phi$ is the azimuthal angle, $D_2$ is the second-order dispersion coefficient (i.e. GVD), $\kappa_\mathrm{r}$ is the microresonator decay rate, $g$ is the single-photon-induced frequency shift due to the Kerr effect, and $\kappa_\mathrm{sc}$ is the backscattering-induced coupling rate between the cw and the ccw modes. 
Since the optical power in the ccw mode is much weaker than the power in the laser-pumped cw mode, and is usually below the parametric four-wave-mixing threshold, here we treat the ccw field as $\phi$-independent, and only the field in the central mode of the microcomb in the frequency domain ($A_\mathrm{cw}^0$) is directly coupled to $A_\mathrm{ccw}$ via the backscattering. 
The coupling rate between the microresonator and the laser is denoted by $\kappa_\mathrm{inj}$, and $\theta$ is the coupling phase determined by the optical feedback path length. 
In order to increase the computation efficiency, $|E_\mathrm{laser}|^2$ is the photon density in the laser cavity for the semiconductor laser system, while $|A_\mathrm{cw}|^2$ and $|A_\mathrm{ccw}|^2$ are the intracavity photon numbers for the Kerr microresonator system. 
The parameter $\eta$ is introduced to the coupling between the laser and the microresonator, so the optical field profiles with different units in the laser and the microresonator are correctly related. 
Table \ref{table1} summarises the parameter values we use for the simulations.

\begin{figure*}[t!]
\renewcommand{\figurename}{\textbf{Supplementary Figure}}
\centering
\includegraphics[width=0.6\textwidth]{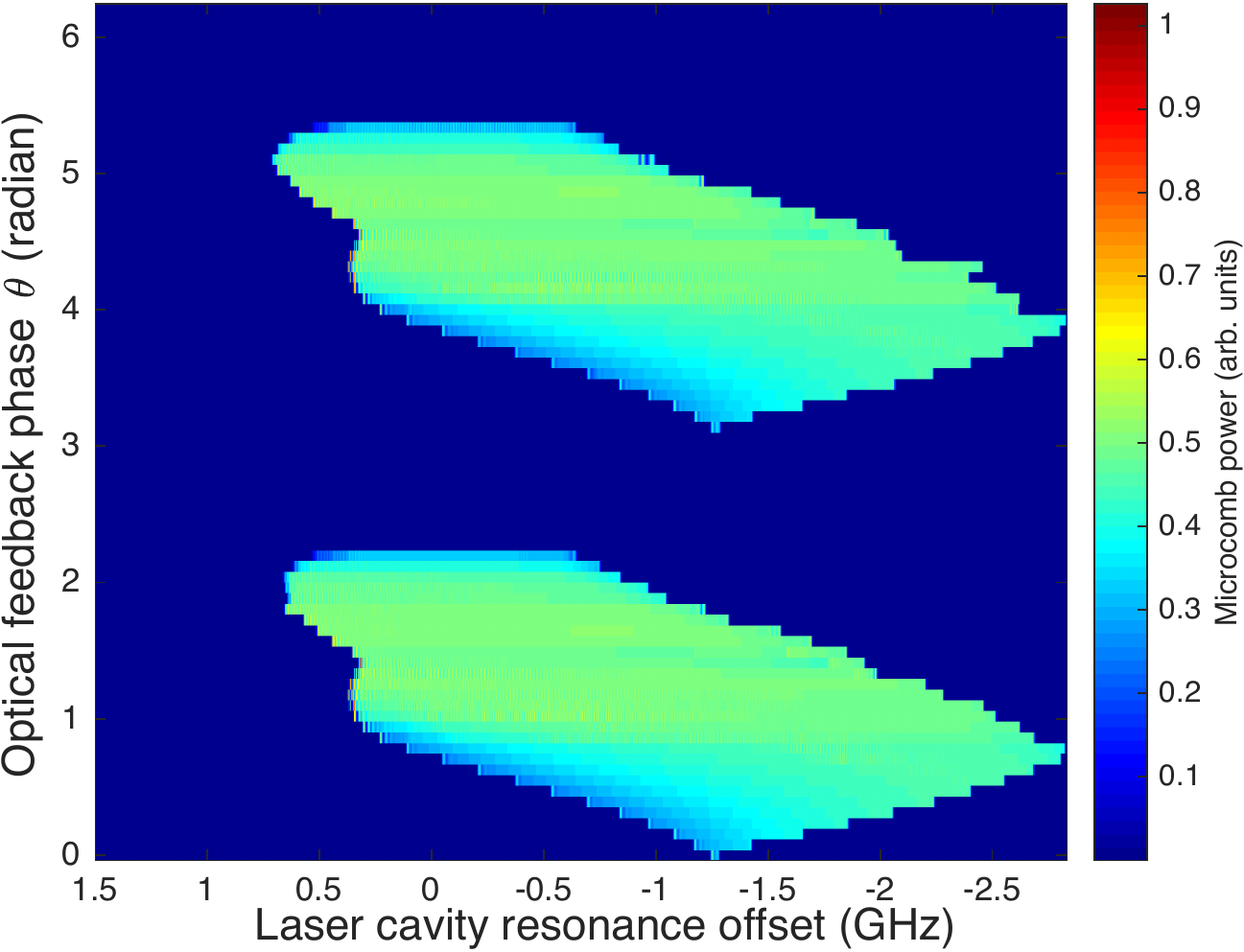} 
\caption{
\textbf{Simulated microcomb power generation with varied optical feedback phase and laser cavity resonance offset.} 
The microcomb power is derived using intracavity power excluding the power in the pumped mode. 
The phase is varied from 0 to $2\pi$ with a step size of $\pi/40$. 
For each feedback phase value, the laser cavity resonance offset is scanned towards a lower frequency to reproduce the experimental conditions.
}\label{figS1}
\end{figure*}

The coupled equations with the optical feedback phase swept from 0 to $2\pi$ are solved in the frequency domain using Runge-Kutta method to yield the temporal evolution of the laser field and the microresonator intracavity fields. 
At each phase value, the laser cavity resonance frequency is frequency-down-swept over a pumped microresonator mode, and complex random noise of very weak magnitude is added to the microresonator intracavity field, such that the comb formation due to modulation instability can be excited when the intracavity power level exceeds the instability threshold. 
Supplementary Figure \ref{figS1} shows the contour plot of the simulated intracavity microcomb power (i.\,e., the intracavity power excluding the power in the pumped mode). 
This result shows that the successful microcomb generation strongly relies on the appropriate choice of the optical feedback phase, and that the comb generation pattern exhibits a $\pi$-periodic dependence on the feedback phase, which are in qualitative agreement with the experimental observations shown in Fig. 4 in the main manuscript).

%%%%%%%%%%%%%%%%%%%%%%%%%%%%%%%%%%
%%%%%%%%%%%%%%%%%%%%%%%%%%%%%%%%%%

In Supplementary Fig. \ref{figS2}, we compare in detail the system dynamics with two different feedback phases ($\theta = 0.2 \pi$ for Supplementary Fig. \ref{figS2} (a, b, c) and $\theta = \pi$ for Supplementary Fig. \ref{figS2} (d, e, f)). 
As can be seen from the laser frequency offset (with regard to the microresonator's cold-cavity-mode resonance frequency) curves, in both situations the laser can be self-injection locked for certain laser-microresonator frequency detuning ranges, and the laser is slightly red-detuned (i.\,e., with a negative lasing frequency offset) in both cases due to the Kerr-nonlinearity-induced microresonator frequency shift. 
As the laser cavity resonance frequency is down-swept, in the first case the intracavity power rises, at certain point accumulating enough energy for microcomb initiation with the help of intracavity noises, while in the latter case the intracavity power level keeps decreasing, thus never reaching the threshold for microcomb generation. 
These behaviours have been previously well studied theoretically and numerically \cite{Kondratiev:17, Voloshin:21}.

%%%%%%%%%%%%%%%%%%%%%%%%%%%%%%%%%%
\begin{figure*}[t!]
\renewcommand{\figurename}{\textbf{Supplementary Figure}}
\centering
\includegraphics[width=0.9\textwidth]{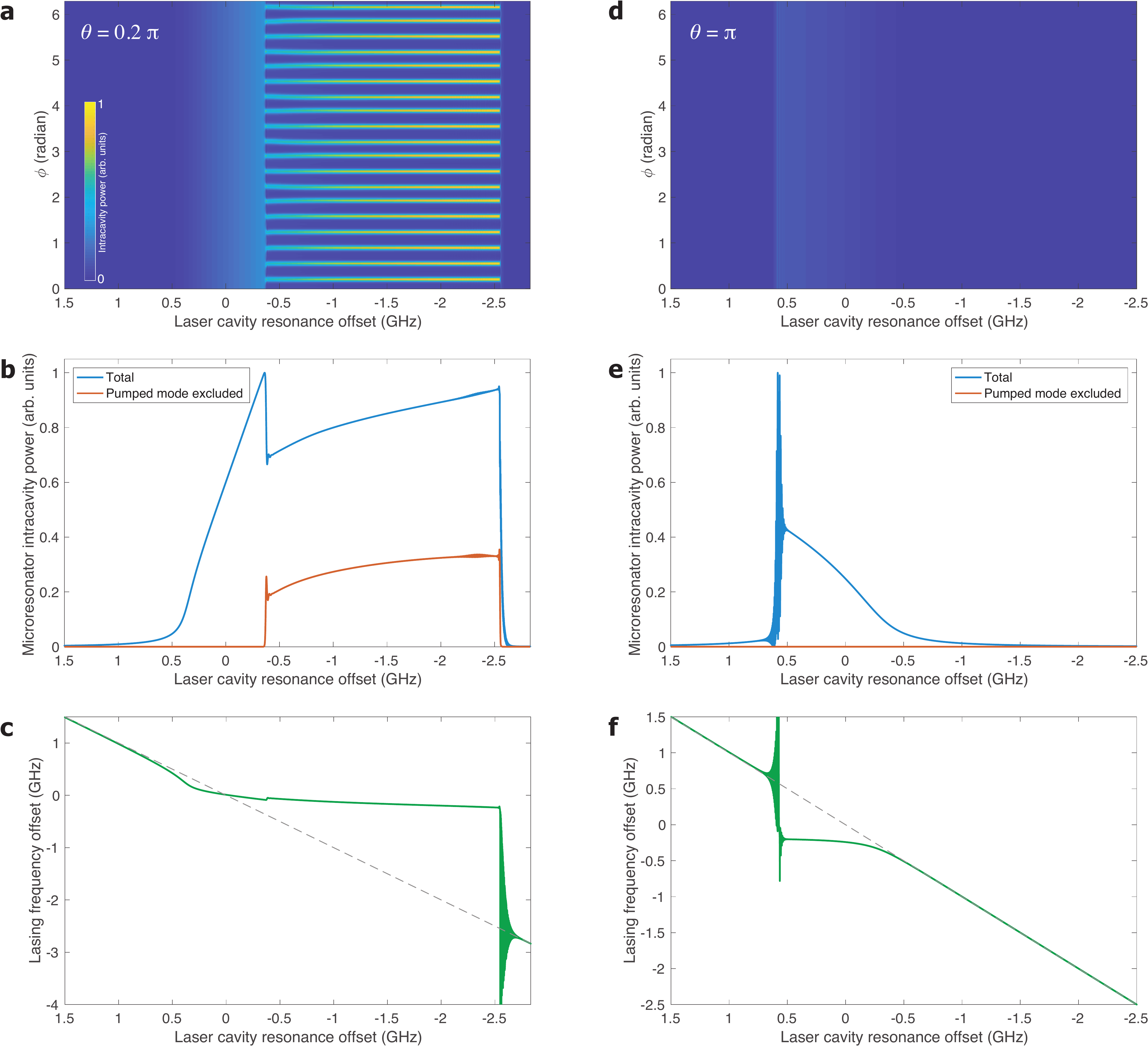} 
\caption{
\textbf{The evolution of microresonator intracavity field and the laser frequency with the feedback phase $\theta = 0.2\pi$ (left column) and $\pi$ (right column), respectively.} 
\textbf{(a, d)} The spatiotemporal intracavity power evolution in the Kerr microresonator. 
\textbf{(b, e)} The normalized intracvity power evolution of the cw direction. 
\textbf{(c, f)} The lasing frequency (with regard to the cold-cavity frequency of the pumped mode in the microresonator) as the laser cavity frequency is frequency-down swept. 
The dashed grey lines indicate the laser frequency change when the feedback is disabled.
}\label{figS2}
\end{figure*}
%%%%%%%%%%%%%%%%%%%%%%%%%%%%%%%%%%

\bibliographystyle{apsrev4-1}
\bibliography{bibliography}